\documentclass[acmsmall,screen]{acmart}
\AtBeginDocument{%
  }

\setcopyright{acmlicensed}
\copyrightyear{2018}
\acmYear{2018}
\acmDOI{XXXXXXX.XXXXXXX}
\acmConference[ACM Transactions of Software Engineering and Methodology]

\usepackage[T1]{fontenc}
\usepackage[utf8]{inputenc}

\usepackage{amsmath,amsfonts}

\usepackage{color,hyperref}
\usepackage{graphicx}
\usepackage[all]{nowidow}
\usepackage{colortbl}
\usepackage{xspace}
\usepackage[all]{nowidow}
\usepackage[inline]{enumitem}
\usepackage[scaled]{beramono}
\usepackage{booktabs} 
\usepackage[colorinlistoftodos]{todonotes}
\usepackage{multirow}
\usepackage{balance}
\usepackage{listings}
\usepackage{microtype}
\usepackage{csquotes}

\usepackage{algpseudocode}

\usepackage{multirow}
\usepackage[many]{tcolorbox}
\usepackage{collectbox}
\usepackage{cleveref}
\usepackage{tikz}
\usepackage{collcell}
\usepackage[utf8]{inputenc}
\usepackage{xstring}
\usetikzlibrary{shadows}
\usepackage{caption}
\usepackage{subcaption}

\definecolor{darkblue}{rgb}{0.1,0.6,1.0}
\definecolor{red}{rgb}{1.0, 0.01, 0.24}
\definecolor{gray}{rgb}{0.8, 0.8, 0.8}
\definecolor{darkgreen}{rgb}{0.0, 0.5, 0.0}
\definecolor{bluebell}{rgb}{0.64, 0.64, 0.82}
\definecolor{bluegray}{rgb}{0.4, 0.6, 0.8}
\hypersetup{colorlinks,breaklinks,
	linkcolor=darkblue,urlcolor=darkblue,
	anchorcolor=darkblue,citecolor=darkblue}

\usepackage{adjustbox}

\usepackage{algorithm}
\usepackage{algpseudocode}
\usepackage{dutchcal}
\usepackage{caption}
\usepackage[many]{tcolorbox}
\usepackage{MnSymbol}
\usepackage{longtable}
\usepackage{comment}

\algblock{Input}{EndInput}
\algnotext{EndInput}
\algblock{Output}{EndOutput}
\algnotext{EndOutput}
\newcommand{\Desc}[2]{\State \makebox[2em][l]{#1}#2}

\newcommand{\rqone}{\textbf{RQ$_1$}: \emph{Does the inclusion of source code help to enhance the detection of technical debt?}} 

\newcommand{\rqtwo}{\textbf{RQ$_2$}: \emph{What is the accuracy of different pre-trained models when detecting TD solely from source code?}} 

\newcommand{\rqthree}{\textbf{RQ$_3$}: \emph{How do the manually classified comments contribute to 
the detection of SATD?}}

\lstdefinelanguage{m3}{
	basicstyle=\ttfamily\scriptsize,
	keywordstyle=\bfseries,
	keywords={m3,declarations,methodInvocation},
	literate={<-}{$\leftarrow$}{1},
	tabsize=2,
	alsoletter={-}
}



\usepackage{xspace}
\newcommand*{\ie}{i.e.,\@\xspace}
\newcommand*{\eg}{e.g.,\@\xspace}
\newcommand*{\etal}{et al.\@\xspace}

\newcommand{\GH}{\textsc{GitHub}\xspace}
\newcommand\revised[1]{\textcolor{black}{#1}}

\makeatletter
\newcommand*{\etc}{%
	\@ifnextchar{.}%
	{etc}%
	{etc.\@\xspace}%
}
\makeatother

\definecolor{verylightgray}{gray}{0.98}

\newtcolorbox{shadedbox}{
	drop shadow southeast,
	breakable,
	enhanced jigsaw,
	colback=white,
	boxrule=0.80pt,
	left=0.3em,
	right=0.3em,
	top=0.1em,
	bottom=0.05em
}

\definecolor{mygreen}{rgb}{0,0.6,0}
\definecolor{mygray}{rgb}{0.95,0.95,0.95}
\definecolor{myred}{rgb}{0.5,0,0}

\lstdefinestyle{JavaStyle} {
	backgroundcolor=\color{verylightgray},   
	commentstyle=\color{mygreen}, 
	breakatwhitespace=false,
	keywordstyle=\color{violet},
	language=Java,
	stringstyle=\color{blue},
	basicstyle=\tiny\ttfamily,
	showstringspaces=false}

\newcommand*{\MinNumber}{0.00}%
\newcommand*{\MidNumber}{50.00} %
\newcommand*{\MaxNumber}{100.00}%
\newcommand*\mynum{}
\definecolor{ao(english)}{rgb}{0.0, 0.5, 0.0}
\definecolor{cadmiumgreen}{rgb}{0.0, 0.42, 0.24}
\definecolor{emerald}{rgb}{0.31, 0.78, 0.47}

\newcommand{\ApplyGradient}[1]{%
	\IfDecimal{#1}{%
		\edef\mynum{#1}%
		\ifdim #1 pt > \MaxNumber pt\relax
		\edef\mynum{\MaxNumber}%
		\else
		\ifdim #1 pt < \MinNumber pt\relax
		\edef\mynum{\MinNumber}%
		\fi
		\fi
		\ifdim \mynum pt > \MidNumber pt
		\pgfmathsetmacro{\PercentColor}{max(min(100.0*(\mynum - \MidNumber)/(\MaxNumber-\MidNumber),100.0),0.00)}%
		\xdef\PercentColorr{\PercentColor}
		\cellcolor{white!\PercentColorr!emerald}#1
		\else
		\pgfmathsetmacro{\PercentColor}{max(min(100.0*(\MidNumber - \mynum)/(\MidNumber-\MinNumber),100.0),0.00)}%
		\xdef\PercentColorr{\PercentColor}
		\cellcolor{white!\PercentColorr!emerald}#1%
		\fi	
	}{\textbf{#1}}
}

\newcolumntype{R}{>{\collectcell\ApplyGradient}c<{\endcollectcell}}

\newcommand*{\TS}{\textsc{Tesoro}\@\xspace}
\newcommand*{\TScomment}{\textsc{Tesoro}$_{comment}$\@\xspace}
\newcommand*{\TScode}{\textsc{Tesoro}$_{code}$\@\xspace}
\newcommand*{\Maldonado}{\texttt{Maldonado-62K}\@\xspace}

\colorlet{punct}{red!60!black}
\definecolor{background}{HTML}{EEEEEE}
\definecolor{delim}{RGB}{20,105,176}
\colorlet{numb}{magenta!60!black}
\lstdefinelanguage{json}{
	basicstyle=\footnotesize\ttfamily,
	numbers=left,
	numberstyle=\scriptsize,
	stepnumber=1,
	numbersep=8pt,
	showstringspaces=false,
	breaklines=true,
	frame=lines,
	backgroundcolor=\color{background},
	literate=
	*{0}{{{\color{numb}0}}}{1}
	{1}{{{\color{numb}1}}}{1}
	{2}{{{\color{numb}2}}}{1}
	{3}{{{\color{numb}3}}}{1}
	{4}{{{\color{numb}4}}}{1}
	{5}{{{\color{numb}5}}}{1}
	{6}{{{\color{numb}6}}}{1}
	{7}{{{\color{numb}7}}}{1}
	{8}{{{\color{numb}8}}}{1}
	{9}{{{\color{numb}9}}}{1}
	{:}{{{\color{punct}{:}}}}{1}
	{,}{{{\color{punct}{,}}}}{1}
	{\{}{{{\color{delim}{\{}}}}{1}
	{\}}{{{\color{delim}{\}}}}}{1}
	{[}{{{\color{delim}{[}}}}{1}
	{]}{{{\color{delim}{]}}}}{1},
}

\makeatletter
\newtcolorbox{mytcbox}[2][]{%
	enhanced, 
	breakable,
	colback=white,
	colframe=bluegray!40!black,
	attach boxed title to top left={yshift=-2pt}, title={#2},
	boxed title size=standard,
	boxrule=0pt,
	boxed title style={%
		sharp corners, 
		rounded corners=northwest, 
		colback=tcbcolframe, 
		boxrule=0pt},
	sharp corners=north,
	overlay unbroken={%
		\path[fill=tcbcolback] 
		([xshift=-2pt]title.south east) 
		to[out=0, in=180] ([xshift=1.5cm]title.east)--
		(title.east-|frame.east) |- 
		([xshift=-2pt]title.south east)--cycle;
		\path[fill=tcbcolframe] (title.south east) 
		to[out=0, in=180] ([xshift=1.5cm]title.east)--
		(title.east-|frame.east)
		[rounded corners=\kvtcb@arc] |- 
		(title.north-|frame.north) 
		[sharp corners] -| (title.south east);
		\draw[line width=.5mm, rounded corners=\kvtcb@arc, 
		tcbcolframe] 
		(title.north west) rectangle 
		(frame.south east);
	}, 
	overlay first={%
		\path[fill=tcbcolback] 
		([xshift=-2pt]title.south east) 
		to[out=0, in=180] ([xshift=1.5cm]title.east)--
		(title.east-|frame.east) |- 
		([xshift=-2pt]title.south east)--cycle;
		\path[fill=tcbcolframe] (title.south east) 
		to[out=0, in=180] ([xshift=1.5cm]title.east)--
		(title.east-|frame.east)
		[rounded corners=\kvtcb@arc] |- 
		(title.north-|frame.north) 
		[sharp corners] -| (title.south east);
		\draw[line width=.5mm, rounded corners=\kvtcb@arc, 
		tcbcolframe] 
		(frame.south west) |- (title.north) -| 
		(frame.south east);
	}, 
	overlay middle={%
		\draw[line width=.5mm, tcbcolframe] 
		(frame.north west)--(frame.south west) 
		(frame.north east)--(frame.south east);
	}, 
	overlay last={%
		\draw[line width=.5mm, rounded corners=\kvtcb@arc, 
		tcbcolframe] 
		(frame.north west) |- (frame.south) -|
		(frame.north east);
	}, 
	#1
}
\makeatother

\begin{document}


\title{\revised{Detection of Technical Debt in Java Source Code}}

\author{Nam Le Hai}
\email{namlh@soict.hust.edu.vn}
\affiliation{%
  \institution{Hanoi University of Science and Technology}
  \city{Hanoi}
  \country{Vietnam}
}

\author{Anh M. T. Bui}
\email{anhbtm@soict.hust.edu.vn}
\affiliation{%
	\institution{Hanoi University of Science and Technology}
	\city{Hanoi}
	\country{Vietnam}
}

\author{Phuong T. Nguyen}


\affiliation{%
	\institution{University of L'Aquila}
	\city{L'Aquila}
	\country{Italy}
}
\email{phuong.nguyen@univaq.it}

\author{Davide Di Ruscio}
\affiliation{%
	\institution{University of L'Aquila}
	\city{L'Aquila}
	\country{Italy}
}
\email{davide.diruscio@univaq.it}

\author{Rick~Kazman}
\affiliation{%
	\institution{University of Hawaii}	
	\city{Hawaii}
	\country{USA}
}
\email{kazman@hawaii.edu}

%
%
%
%
%
%
%
\renewcommand{\shortauthors}{Le et al.}

\begin{abstract}
Technical debt (TD) describes 
the additional 
costs that emerge when developers have opted for a quick and easy solution to a problem,  rather than a more effective and well-designed, but time-consuming approach. 
Self-Admitted Technical Debts (SATDs) are a specific type of technical debts that developers intentionally document and acknowledge, typically 
via textual comments. 
While these 
comments are a useful tool for identifying TD, 
most of the existing approaches 
focus on capturing 
tokens associated with various categories of TD, neglecting the rich information embedded within the source code. 
Recent research has focused on detecting SATDs by analyzing comments, 
and there has been little work dealing with TD 
contained in the source code. 
In this study, through the analysis of comments and their 
source code from 974 Java projects, 
we curated 
the first ever dataset of TD identified by code comments, coupled with its 
code. %
We found that
including the classified 
code significantly improves the accuracy in predicting various types of technical debt. 
We believe that our dataset will catalyze future work in the domain, inspiring various research 
related to the recognition of technical debt; 
The proposed classifiers may serve as baselines for 
studies on the detection of TD. 

\end{abstract}



\maketitle

\section{Introduction}
\label{sec:introduction}

The concept of technical debt was originally introduced by Cunningham~\cite{cunningham1992wycash} to represent the liabilities that arise when developers make sub-optimal technical decisions, either intentionally or unintentionally during the software development life-cycle in their rush to market. Various factors can lead to the accumulation of technical debt, including deadline pressures, existing low-quality code, misaligned incentives, and poor software processes, among others~\cite{bellomo2016got}. Previous studies have shown that developers often underestimate the consequences of such debts, which can degrade the quality of the source code, increase bug rates, and slow development velocity~\cite{zazworka2011investigating,wehaibi2016examining}. Identifying the code that contains technical debt is crucial to a rational development process, as this allows developers to fix the most important issues, the ones that are slowing the project down.

In 2015, da Silva Maldonado \etal \cite{maldonado2015detecting} introduced their seminal work on detecting self-admitted technical debt (SATD) from comments embedded in source code. In particular, the authors manually classified a set of code comments to identify various types of technical debt. Afterward, they also developed various models to detect SATD from their dataset
~\cite{da2017using}. Significant research on the recognition of SATD has flourished since then~\cite{9804499,sala2021debthunter}.  Recently, there have been various approaches proposed to recognize technical debt contained in different project artifacts. 
Among others, Li \etal~\cite{li2023automatic} conceived an approach to identify SATD from four independent sources, \ie source code comments, commit messages, pull requests, and issue tracking systems. Tan \etal~\cite{TAN2023107216} manually 
curated a dataset of TD manifested in 3,000 issues. An evaluation of the collected dataset showed that there is a positive correlation between the number of TD items identified and mentioned as resolved in issue trackers and the number of debt items paid back in the source code.




The majority of research thus far has mined TD 
from textual sources, \eg 
comments \cite{maldonado2015detecting}, issues~\cite{TAN2023107216,li2023automatic}, or pull requests~\cite{li2023automatic}. In this respect, SATD detection tools rely heavily on text 
to function. 
If no technical debt is reported in text, 
a debt might exist, but these tools would fail to identify it. To add insult to injury, even when there are 
comments, many of them and the corresponding code are not coherent, \ie comments may be out of date and incorrectly reflect what is actually contained in the associated code~\cite{DBLP:journals/sqj/CorazzaMS18}. This may happen because developers forget to update their comments after they made changes to the code~\cite{10.1145/3661167.3661187, rabbi2020detecting}. %

\revised{There have been various attempts to associate SATD with weaknesses. Russo \etal \cite{10.1145/3524842.3528469} conceived WeakSATD 
to analyze source code written in C contained in Chromium projects to understand whether code blocks associated with SATD comments 
may contain weaknesses. The authors curated a set of heuristics from the public Common Weaknesses and Enumeration (CWE) repository to detect known weaknesses in software code and recommend mitigations. Other authors have investigated the possibility of detecting technical debt directly in source code. Nevertheless, these studies typically focused on limited classification schemes--such as distinguishing only between high and not high TD \citep{10586898, 9622154}, or identifying the mere presence or absence of code smells \citep{yadav2024machine}, without considering the broader diversity of technical debt types. 
To the best of our knowledge, no work has been conducted to create large-scale datasets of technical debt directly contained in Java source code. But we see a need for this type of data, to reduce the dependence on textual comments in detecting TD, 
thus vastly enhancing the contexts in which debt may be detected.} 

\revised{To address these challenges, we conceive a new way of detecting technical debt. In particular,} we propose a pipeline for the enrichment 
of technical debt data. Our methodology involves extracting 
SATD comments in conjunction with corresponding source code units. We have devised a method to identify Java source code that possibly contains technical debt, creating our initial corpus. Then we manually classified five categories of technical debt in this corpus. In addition, 
we developed a machine learning based tool to detect technical debt contained in textual comments and source code. By means of an empirical evaluation, %
we demonstrated that the curated dataset has the potential to advance state-of-the-art research in the domain, paving the way for a completely new way of identifying TD. 

Using this dataset, we have addressed the following research questions (RQs):

\begin{itemize}


    \item \rqone~We enriched the input data 
    with source code and fine-tuned four machine learning models, \ie BERT, RoBERTa, UniXCoder, CodeBERT to identify technical debt in code. This RQ aims to investigate whether the enriched dataset (consisting of classified comments and corresponding source code) is beneficial to the detection of technical debt. 
    
    \item \rqtwo~Using various machine learning models, we ran experiments on the collected dataset to investigate the accuracy of these models in classifying 
    debt contained solely in source code. With this RQ we investigate to what extent existing deep learning algorithms are able to detect TD from 
    code, thus inspiring future research in this direction.

        
    \item \rqthree~We augmented the dataset collected by da Silva Maldonado \etal~\cite{maldonado2015detecting} with our newly classified comments to yield a combined dataset. Afterward, we ran four machine learning models, \ie BERT, RoBERTa, UniXCoder, CodeBERT, on both datasets to compare the prediction performance of these models. This aims to determine whether the new comments are useful in predicting technical debt. 
    
        
    
    
    
\end{itemize}


\vspace{.2cm}
\noindent
\textbf{Contributions.} 
In summary, our paper makes the following contributions:
\begin{itemize}
    \item  
    A comprehensive pipeline from data extraction to labeling, aimed at improving labeling efficiency by selecting informative examples to enrich the existing corpus. Given sufficient resources and computational capabilities, this pipeline can be iteratively executed to continuously improve the quality of the dataset.
    
    \item A dataset--named \TS--curated for detecting \underline{\textbf{te}}chnical debt within \underline{\textbf{so}}u\underline{\textbf{r}}ce c\underline{\textbf{o}}de. 
    In addition to existing corpora, \TS offers an additional and important feature, \ie source code that 
    contains debt. 
    This facilitates the exploration of a broader range of scenarios to advance the detection of technical debt. 

    \item We propose novel approaches that integrate source code information to enhance SATD detection. Additionally, we conduct a comprehensive study on effectively utilizing this context by examining the impact of code context length provided to the model.

    \item  
    An empirical study on the curated dataset to evaluate the extent to which it contributes to the detection of technical debt contained in source code. With this evaluation, we attempt to lay the foundations of a new method to identify technical debt, focusing on debts that exist in source code.

    \item The replication package including the curated dataset and the source code implementation has been published online to foster future research.\footnote{\url{https://github.com/NamCyan/tesoro}}

\end{itemize}

\noindent
\textbf{Structure.} 
Section~\ref{sec:Background} provides some background on 
technical debt, as well as the related work. 
Afterward, Section~\ref{sec:Approach} presents in detail the proposed pipeline to curate the dataset. Section~\ref{sec:Datasets} describes the resulting datasets. In Section~\ref{sec:Experiment} we present an empirical study on the usage of 
the resulting dataset to evaluate its effect 
in the detection of technical debt. Section~\ref{sec:Discussion} provides some discussions on the findings, as well as highlights the threats to validity of the results. In Section~\ref{sec:RelatedWork}, we review the related work on the detection of technical debt from different types of input data. Finally, Section~\ref{sec:Conclusions} sketches future work and concludes the paper.

\section{Background}
\label{sec:Background}
In this section, we review different types of SATD and provide an overview of pretrained language models.

\subsection{Self-admitted technical debt (SATD)}


SATD is the technical debt that is expressly admitted by a developer through comments embedded in source code, issue trackers~\cite{9226330}, commit messages, or pull requests~\cite{li2023automatic}. 
da Silva Maldonado \etal \cite{maldonado2015detecting} identified five types of SATD, \ie DESIGN, DEFECT, DOCUMENTATION, REQUIREMENT/IMPLEMENTATION, and TESTING. 
This categorization allows for more insightful descriptions and a deeper understanding of the non-optimal solution options taken. This section takes various examples to explain these SATD categories.






\vspace{.2cm}
\noindent
\textbf{DESIGN.} Comments of this type indicate that there is a problem with the design of the code, \ie 
comments about misplaced code, lack of abstraction, long methods, poor implementation, workarounds, or temporary solutions. To illustrate, consider the following examples.

\begin{tcolorbox}[boxrule=0.86pt,left=0.3em, right=0.3em,top=0.1em, bottom=0.05em]
\small{\textbf{C$_1$}: ``// TODO: - This method is too complex, lets break it up''} \\

\small{\textbf{C$_2$}: ``// I hate this so much even before I start writing it. // Re-initializing a global in a place where no-one will see it just // feels wrong. Oh well, here goes.''} \\

\small{\textbf{C$_3$}: ``//quick \& dirty, to make nested mapped p-sets work:''} \\

\small{\textbf{C$_4$}: ``//I can’t get my head around this; is encoding treatment needed here?''}

\end{tcolorbox}


In \textbf{C$_1$}, the developer said that the method is complex and should be broken up. This is related to the existing design, and the creator of the code signaled this so that other developers could tackle the issue later on. %
In \textbf{C$_2$} the developer complained about the fact that re-initializing a global variable in an obscure location is not the right thing to do. This is actually a design issue, and it needs to be fixed. %
\textbf{C$_3$} implies that the code is a makeshift solution, \ie it is suboptimal, but still works in the given context. 
And in \textbf{C$_4$} the developer wondered aloud if the encoding treatment was really necessary.

\vspace{.2cm}
\noindent
\textbf{DEFECT.} In this category, the authors state that a part of the code does not have the expected behavior, \ie 
there is a lingering defect in the code as shown in the examples below.

\begin{tcolorbox}[boxrule=0.86pt,left=0.3em, right=0.3em,top=0.1em, bottom=0.05em]
\small{\textbf{C$_5$}: ``// Bug in above method''} \\

\small{\textbf{C$_6$}: ``// WARNING: the OutputStream version of this doesn’t work!''} \\

\small{\textbf{C$_7$}: ``// the following stuff did not work and I don't know why!''} \\

\small{\textbf{C$_8$}: ``// POTENTIAL FLAW: Use password directly in PasswordAuthentication()''}

\end{tcolorbox}

\textbf{C$_5$} explicitly points out that there is a bug detected in the given method. This is a clear case of a defect, and it should be fixed soon. With \textbf{C$_6$}, a warning is given, marking OutputStream as a malfunctioning API call in the current context. In \textbf{C$_7$} the developer warned that the code did not work, thereby admitting that they had no idea why this had happened. Eventually, \textbf{C$_8$} signals a disclosure of sensitive information, which possibly poses a security threat. 

\vspace{.2cm}
\noindent
\textbf{DOCUMENTATION.} In this type of debt 
authors express that there is no proper documentation supporting some part of the system. We consider the following examples.

\begin{tcolorbox}[boxrule=0.86pt,left=0.3em, right=0.3em,top=0.1em, bottom=0.05em]
\small{\textbf{C$_9$}: ``// FIXME This function needs documentation''} \\

\small{\textbf{C$_{10}$}: ``// TODO Document the reason for this''} \\

\small{\textbf{C$_{11}$}: ``// @return DOCUMENT ME!''} \\

\small{\textbf{C$_{12}$}: ``// TODO(saurabh): Explain reload scenario here''}
        
\end{tcolorbox}

All the four comments, \ie \textbf{C$_9$}, \textbf{C$_{10}$}, \textbf{C$_{11}$}, and \textbf{C$_{12}$} clearly state that documentation is needed in the containing projects; \textbf{C$_{12}$} is more specific, stressing that it is necessary to explain a concrete method.

\vspace{.2cm}
\noindent
\textbf{REQUIREMENT or IMPLEMENTATION.} Requirement or implementation debt comments express incompleteness of the functionality in the method, class, or program. Here are some examples.

\begin{tcolorbox}[boxrule=0.86pt,left=0.3em, right=0.3em,top=0.1em, bottom=0.05em]
\small{\textbf{C$_{13}$}: ``//TODO no methods yet for getClassname''} \\

\small{\textbf{C$_{14}$}: ``//TODO no method for newInstance using a reverse-classloader''} \\

\small{\textbf{C$_{15}$}: ``/*TODO: The copy function is not yet * completely implemented - so we will * have some exceptions here and there.*/''} \\

\small{\textbf{C$_{16}$}: ``//TODO Find a way to re-send the message.''}

\end{tcolorbox}

Starting with ``//TODO'', \textbf{C$_{13}$} signals the missing implementation for getClassname. Similarly, \textbf{C$_{14}$} indicates the case where the newInstance method is incomplete. In \textbf{C$_{15}$}, the developer admitted that the copy function had not been fully implemented, and will throw some exceptions. \textbf{C$_{16}$} advises developers to look for a suitable method to re-send the messages.

\vspace{.2cm}
\noindent
\textbf{TESTING.} These comments signal the need for the creation or improvement of the current set of tests.

\begin{tcolorbox}[boxrule=0.86pt,left=0.3em, right=0.3em,top=0.1em, bottom=0.05em]

\small{\textbf{C$_{17}$}: ``// TODO - need a lot more tests''} \\

\small{\textbf{C$_{18}$}: ``// TODO enable some proper tests!!''} \\

\small{\textbf{C$_{19}$}: ``// TODO(lwhite): Better tests''} \\

\small{\textbf{C$_{20}$}: ``// TODO figure out how to test this.''}

\end{tcolorbox}

All the 
examples in this category 
indicate that some project members knew that these areas of the code were inadequately tested. Especially, by \textbf{C$_{20}$}, it is highly probable that the developers had not tested the code at all.

In this work, 
we utilized SATD comments as a means to locate source code that possibly contains technical debt.

\subsection{Pretrained Language Models}
\label{sec:lm}

Language models (LMs) are a foundational component in natural language processing (NLP) that have significantly advanced over the past decade. 
Recently, LMs have been powered by neural networks and trained on large text corpora, being able 
to capture both the syntactic and semantic aspects of languages more effectively. These models commonly follow the pre-training and fine-tuning paradigm \citep{zhao2023survey}. During the pre-training phase, models are trained on large-scale unlabeled corpora using task-agnostic objectives such as word prediction, resulting in the development of pre-trained language models (PLMs). PLMs are then fine-tuned to adapt to various downstream tasks. Early PLMs \citep{sarzynska2021detecting} were mostly based on Recurrent Neural Networks (RNNs) and their variants, such as long short-term memory (LSTM) \cite{le2014sequence} and gated recurrent units (GRU) \cite{cho2014properties}.  However, these approaches were computationally inefficient due to limitations in parallel processing, reducing scalability when training with extensive datasets and large model sizes.  With the introduction of the Transformer architecture \cite{vaswani2017attention} and its self-attention mechanism, significantly more parallelization became possible as compared to RNNs. This advancement enables efficient pre-training of large language models on extensive datasets using multiple GPUs.  Various transformer-based PLMs have achieved state-of-the-art performance across a wide range of tasks \cite{devlin2018bert, liu2019roberta, lan2019albert, raffel2020exploring, touvron2023llama}. Given the superior performance of transformer-based PLMs, which have also been explored in the context of SATD detection~\cite{9804499,sharma2022self,sheikhaei2024empirical}, our study concentrates on utilizing these models.

\subsubsection{Transformer architecture}

This section provides an overview of the Transformer architecture, emphasizing key components and elements \cite{vaswani2017attention}. 

\vspace{.2cm}
\noindent \textit{\textbf{Encoder-Decoder architecture:}} The Transformer architecture, initially designed for machine translation problems, features both an encoder and a decoder. The encoder consists of a stack of six identical layers, each containing two sub-layers: a multi-head self-attention and a position-wise feed-forward neural network.  Similarly, the decoder is structured with six identical layers, but in addition to the two sub-layers found in the encoder, it includes a third sub-layer that applies multi-head attention over the encoder's output. In addition, the decoder uses a masked matrix in the attention layer to prevent attending to future positions in the input sequence, ensuring that the model only considers previously generated tokens during training. 

\vspace{.2cm}
\noindent \textit{\textbf{Multi-head self-attention mechanism:}} The attention mechanism operates by mapping a query and a collection of key-value pairs to an output. The output is obtained by computing a weighted sum of the values, with the weights (or attention scores) derived from a compatibility function that measures the alignment between the query and each corresponding key. Instead of utilizing a single attention mechanism with keys, values, and queries of dimensionality $d_{model}$, it has been found advantageous to project the queries, keys, and values into dimensions $d_q$, $d_k$ and $d_v$, respectively, through distinct learned linear projections (multi-head).

\vspace{.2cm}
\noindent \textit{\textbf{Positional encoding:}} This technique is introduced to integrate information regarding the relative or absolute positions of tokens within the sequence. Specifically, the Transformer model employs absolute positional encoding by utilizing sine and cosine functions to represent token positions.

\subsubsection{Types of PLMs}
\label{sec:plms_type}

Based on the neural architectures of Transformer-based PLMs, we categorize the models into three main groups, as also outlined in existing work \cite{minaee2024large}.

\vspace{.2cm}
\noindent \textit{\textbf{Encoder-based PLMs:}} This type of model utilizes the Transformer Encoder and builds a network by stacking multiple layers. These models were initially developed for language understanding tasks, such as text classification, where the objective is to predict a class label for a given input text. The pre-training stage of these models typically involves corrupting a given sentence in some way (e.g., by masking random words) and then training the model to identify or reconstruct the original sentence. To tackle a downstream task such as sentence classification or named entity recognition, these models are fine-tuned on task-specific data, and this involves substituting the LM head (the word prediction layer), with a classification head. BERT \cite{devlin2018bert}, a prominent encoder-based model, has inspired the development of several variants, such as RoBERTa \cite{liu2019roberta} and ALBERT \cite{lan2019albert}, which have demonstrated substantial improvements across various understanding tasks.

Bidirectional Encoder Representations from Transformers (BERT) \cite{devlin2018bert} is among the most widely adopted encoder-based PLMs. During pretraining, BERT leverages two objectives: masked language modeling (MLM) and next sentence prediction (NSP). In MLM, random tokens within a sentence are masked, and the model is trained to predict these masked tokens using the context of the surrounding words. Meanwhile, NSP trains BERT to comprehend the relationship between two sentences by predicting whether one sentence logically follows the other. RoBERTa \cite{liu2019roberta} extends BERT by improving its robustness through refined model design choices and training strategies. These enhancements include adjusting some key hyperparameters, eliminating the NSP objective, and training with a larger batch size and learning rate. ALBERT \cite{lan2019albert} introduces two parameter reduction techniques to reduce memory consumption and enhance the training speed of BERT.

\vspace{.2cm}
\noindent \textit{\textbf{Encoder-Decoder-based PLMs:}} This neural architecture is primarily designed for sequence-to-sequence tasks, including machine translation, text summarization, and dialogue generation. These models integrate both the encoder and decoder modules of the Transformer, where the encoder processes the input sequence into continuous representations that capture contextual information, and the decoder sequentially generates the output sequence based on these representations. T5 \cite{raffel2020exploring} and BART \cite{lewis2019bart} are two prominent Encoder-Decoder-based PLMs that have demonstrated exceptional performance in sequence-to-sequence tasks.

The Text-to-Text Transfer Transformer (T5) model \cite{raffel2020exploring} advances the field of transfer learning in NLP by proposing a unified framework that reformulates all text-based language tasks into a text-to-text format. BART \cite{lewis2019bart} utilizes a standard sequence-to-sequence model architecture augmented with a denoising strategy, in which the input text is intentionally corrupted using various noising functions such as token masking, document rotation, or sentence permutation. The model is then trained to reconstruct the original text from the corrupted input.

\vspace{.2cm}
\noindent \textit{\textbf{Decoder-based PLMs:}} In these models, the attention layers at each stage are restricted to attending only to preceding words in the sentence, characterizing them as unidirectional or auto-regressive models. The pre-training process generally involves predicting the next word (or token) in the sequence. Consequently, decoder-based models are particularly effective for text generation tasks. The GPT \cite{radford2018improving, radford2019language, brown2020language} and LLaMA \cite{touvron2023llama, touvron2023llama2} families have developed several powerful foundational models that utilize the Transformer’s decoder architecture. These models are pre-trained on extensive datasets comprising trillions of tokens and enhance the architecture through various techniques, such as employing the SwiGLU activation function instead of ReLU, incorporating rotary positional embeddings in place of absolute positional embeddings, and utilizing root-mean-squared layer normalization instead of the standard layer normalization.

\textit{Large Language Models} (LLMs) primarily refer to Transformer-based PLMs characterized by their extensive architecture, containing billions of parameters. These models are primarily inspired by decoder-based architectures, forming the foundation for the development of more advanced LLMs. LLMs are considerably larger in size, exhibiting superior language understanding and generation capabilities compared to small-scale PLMs. Some notable LLMs include GPT-4 \cite{achiam2023gpt}, LLaMA-2 \cite{touvron2023llama2}, PaLM \cite{chowdhery2023palm}, and FLAN \cite{wei2021finetuned}.

\vspace{.2cm}
Based on the data utilized for the pre-training stage, we categorize PLMs into two groups.

\begin{figure*}[!ht]
\centering
\small
\includegraphics[width=0.95\textwidth]{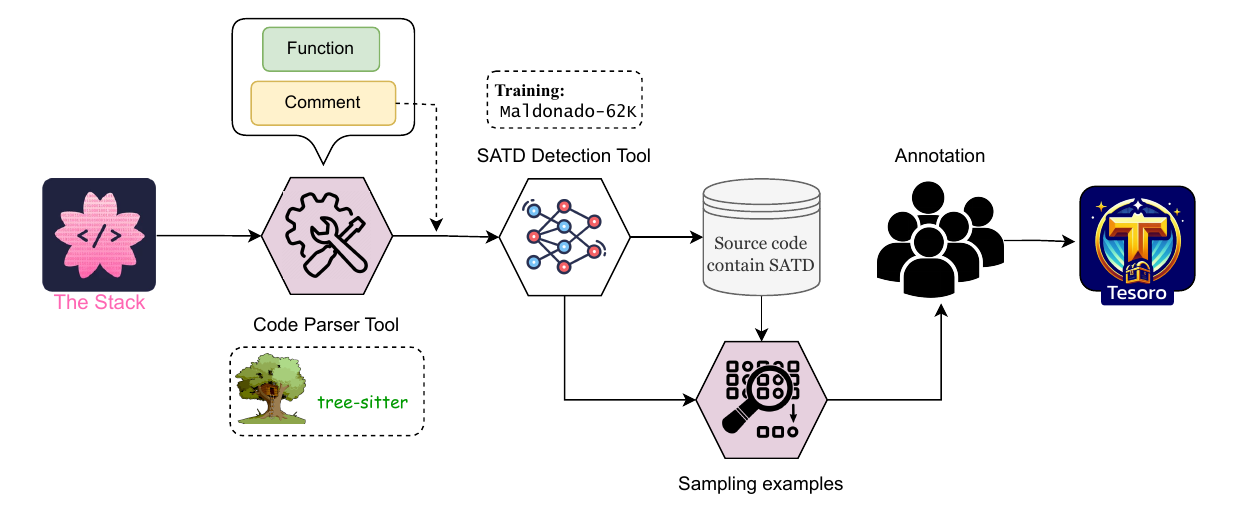}
\caption{An Overview of the \TS Creation Pipeline.}
\label{fig:data_pipeline}
\end{figure*}

\begin{itemize}
 

\item \textit{\textbf{NL-based PLMs:}} This is a class of models primarily trained on extensive natural language text corpora \cite{devlin2018bert, liu2019roberta, lan2019albert, raffel2020exploring, lewis2019bart, radford2018improving, radford2019language, brown2020language, touvron2023llama, touvron2023llama2}. These models leverage vast amounts of textual data to learn rich linguistic representations, making them highly effective for a wide range of NLP tasks, such as text classification, sentiment analysis, and question answering.

\item \textit{\textbf{Code-based PLMs:}} These are specialized models designed to understand and generate programming code \cite{DBLP:conf/emnlp/FengGTDFGS0LJZ20, DBLP:conf/acl/GuoLDW0022, guo2020graphcodebert, nijkamp2022codegen, wang2023codet5+, wang2021codet5, roziere2023code, lozhkov2024starcoder, wei2023magicoder}. Typically, these models are initialized from NL-based PLMs and further trained on large corpora of source code from various programming languages. The datasets are collected from rich code sources, including GitHub and Stack Overflow. These models demonstrate exceptional performance across various code-related tasks, including code summarization, code translation, bug detection, technical debt detection, and code generation.

\end{itemize}

\section{Proposed Methodology}
\label{sec:Approach}

In this section, we describe our proposed approach to constructing the \TS dataset. We outline the steps of our processing pipeline as follows.

\begin{itemize}
    \item We reused the benchmark dataset proposed by Maldonado \textit{et al.}~\cite{maldonado2015detecting} to train an SATD classifier.
    

    \item The pre-trained SATD classifier was then employed to detect SATD comments from open-source projects within the Stack corpus~\cite{kocetkov2022stack}.

    \item We then created an approach to localize and annotate code snippets following SATD comments from open-source projects in the Stack corpus.  We detail the annotation process at the end of this section. Specifically, we invited seven Master's students of Computer Science to verify the SATD comments and their associated source code snippets. The objective was to assign a technical debt (TD) label to the source code. 
    
\end{itemize}

\TS facilitates the detection of technical debt (TD) not only in comments but also through an additional feature: the source code.
For SATD tasks, \TS offers additional code context, rather than relying solely on comments as in previous studies \cite{maldonado2015detecting, da2017using, sharma2022self}. To construct the dataset for TD detection in source code, we employed information from SATD comments to identify specific categories of debt within the code. While in prior work the identification of TD was based solely on comments, in this work we identify where TD appears in the source code, without accompanying comments. 

The data collection process  
is illustrated in Figure \ref{fig:data_pipeline}. The pipeline consists of four major components:  a \textbf{Code Parser Tool} to extract functions and comments from Java files, an \textbf{SATD Detection Tool} to identify TD types in comments, a \textbf{Sampling Strategy} to select high-quality samples, and an \textbf{Annotation Process} to assign a TD type to chosen comments.

\subsection{Source Data}
We initially opted for \GH as our data retrieval source. However, due to rate limit constraints of the \GH API\footnote{\url{https://docs.github.com/en/rest?apiVersion=2022-11-28}}, we adopted an alternative dataset: The Stack \cite{kocetkov2022stack}, which has been acknowledged as the most extensive publicly available source code dataset, boasting a permissive license and a substantial size of 3TB. The Stack contains 
samples from 358 programming languages. In this work, we focus on detecting TD in Java source code. This subset of The Stack yielded a dataset of 26M raw files. Due to constraints in storage and computational resources associated with the code parser and SATD Detection tools, we restricted ourselves to analyzing just 2M of these files. Table~\ref{tab:Annotation_scores} shows the statistics for the dataset across each phase.


When considering identifying TD at the file level, large files present challenges for developers in localizing the sections of code harboring TD. Our goal, then, was to reduce this scope, focusing on identifying TD within the context of function blocks in the source code, as depicted in Figure \ref{fig:extraction}.

\subsection{Code Parser Tool}
\label{sec:parser_tool}

As our emphasis is on detecting TD at the function level, we needed to parse code files from The Stack into individual functions. Since comments serve as the primary annotations for identifying TD within a function, we extracted a collection of functions from each file, each containing a set of comments. We leverage Toolkit\footnote{\url{https://github.com/FSoft-AI4Code/CodeText-parser}}---a tool introduced in our previous work \cite{DBLP:conf/emnlp/ManhHDNNGB23}, which relies on Tree-sitter\footnote{\url{https://tree-sitter.github.io/tree-sitter/}}---to parse source code into Abstract Syntax Tree (AST) representations, enabling the extraction of functions. Subsequently, we extracted a series of comments associated with each function.

\begin{figure*}[t]
	\centering
	\small
	\includegraphics[width=0.92\textwidth]{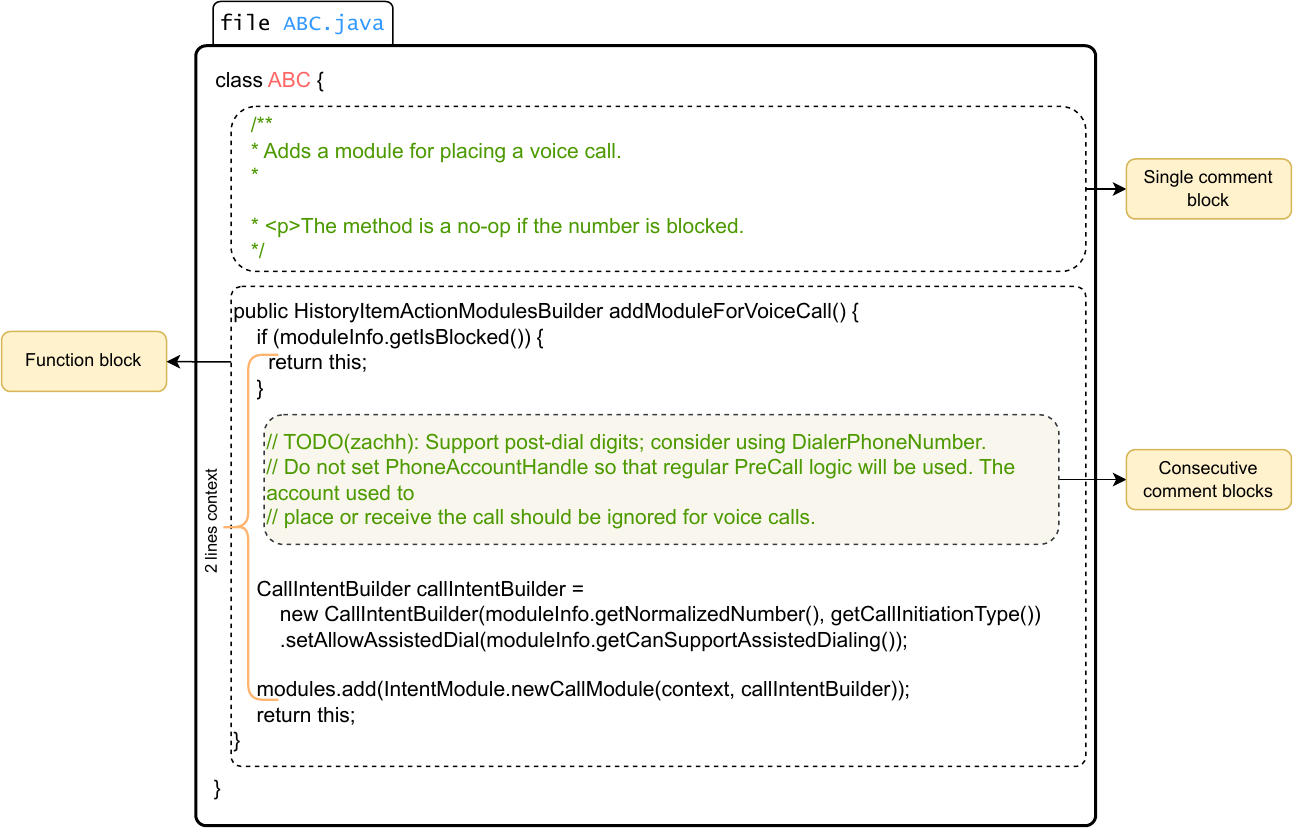}
	\caption{Extraction of comments and functions.}
	\label{fig:extraction}
\end{figure*}

Comments belonging to a function are defined as those located within the body of the function, and the initial comment preceding the function's definition. Developers commonly break down long comments into several lines. The AST classifies distinct lines of comments as individual block nodes; we re-classify consecutive comment lines as a single comment. 
For example, in Figure \ref{fig:extraction}, there are four blocks of comments within the $\mathtt{addModuleForVoiceCall}$ function: three blocks contained within the function and one outside (highlighted in green). The three blocks within the function are consecutive, so we group them to form a single comment. As a result, the function $\mathtt{addModuleForVoiceCall}$ contains two comment statements. If the single comment block in Fig. \ref{fig:extraction} includes a TD, there there are two different data points.



The comments extracted from each function are then annotated following previous research on SATD detection \citep{maldonado2015detecting, da2017using, sierra2019survey, li2023automatic}. 
The labeled information of comments then serves as the ground truth for assigning TD in the function after removing all comments.

\subsection{SATD Detection Tool}
\label{sec:detection_tool}

Since there is a large number of functions and comments, the annotation process requires considerable human labor. 
Thus, 
it becomes crucial to choose a subset for annotation purposes. Previous studies \citep{maldonado2015detecting, da2017using} showed that the majority of extracted comments do not include TD, with over 90\% of comments \textit{not} implying TD. Therefore, randomly selecting examples for annotation might yield numerous comments that do not contribute to the TD identification process.

\textbf{Identify comments containing TD}: To address this challenge, we developed a TD detection tool to identify comments containing TD within the corpus from Section \ref{sec:parser_tool}. 
In particular, we constructed a neural model to determine whether a comment contains TD or not. In fact, there have been various SATD detection techniques~\cite{9804499,sala2021debthunter}, but we decided to develop a tailored tool on top of pre-trained models as a means to validate thei effectiveness in detecting SATD. %
The detection tool is a binary classifier, in which all comments containing TD  are classified into the positive class, while the rest are assigned to the negative class. Since comments are predominantly in natural language text format, we built the tool using the RoBERTa architecture~\cite{liu2019roberta}. 
Since the tool needed to process a substantial volume of comments, we employed the base version of the model, with 128 million parameters, to balance performance and speed. The \Maldonado \cite{da2017using} dataset was leveraged for fine-tuning. The dataset includes more than 62,000 comments, of which 6.5\% were identified as containing debt, and categorized into 5 different types of TD, 
and the majority were identified as non-SATD. We grouped the five classes into the positive class to train the neural classifier using binary cross-entropy loss.

Before fine-tuning, we performed a simple cleaning process to convert 
comments to lowercase, removing comment delimiters such as ``$//$'', ``$\textbackslash$*'', ``*$/$'', and eliminating duplicates. We conducted training for 10 epochs with a learning rate set to $2e-5$, and held back 
10\% of the data as a validation set. Subsequently, the trained model is employed to scan through approximately 40 million comments to seek out those containing TD. Consequently, over 1.6 million comments were identified as potentially implying TD.

\begin{figure}[t]
\centering
\begin{minipage}{0.48\linewidth}

\captionof{table}{Input data information across phases. \label{tab:Annotation_scores}}
\begin{adjustbox}{width=\textwidth}
\begin{tabular}{l|c|c|c}
\toprule
Phase & \#File & \#Function & \#Comment \\
\midrule
Raw files (The Stack-Java) & 26M & - & - \\
Code Parser Tool & 2M & - & - \\
SATD Detection Tool & - & - & 3.6M \\
Annotation process & 999 & 1,255 & 4.981 \\
\bottomrule
\end{tabular}
\end{adjustbox}

\vspace{0.5cm}

\captionof{table}{Annotation Assessment. \label{tab:annotation_scores}}
\begin{adjustbox}{width=\textwidth}
\begin{tabular}{l|c|c|c}
\toprule
Phase & Number of comments & Raw Agreement & IAA \\
\midrule
1 & 1,400 & 56.18 & 37.00 \\
2 & 3,680 & 92.77 & 45.29 \\
\bottomrule
\end{tabular}
\end{adjustbox}

\end{minipage}
\hfill
\begin{minipage}{0.48\linewidth}
\includegraphics[width=0.85\textwidth]{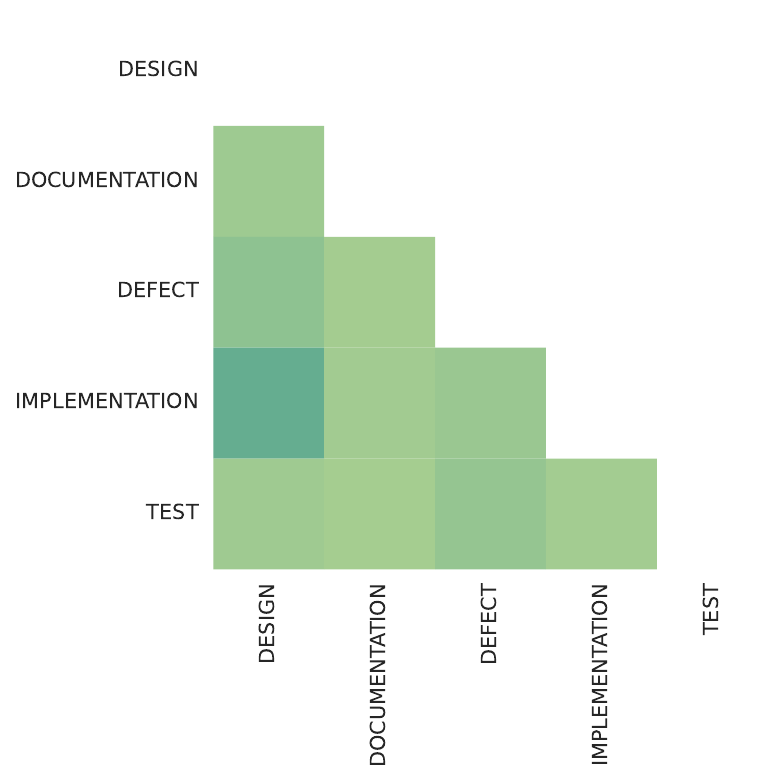}
\caption{Overlap categories ratio from multiple binary classifiers prediction on a comment.}
\label{fig:overlap_category}
\end{minipage}
\end{figure}

\textbf{Detecting TD types of comments}: After acquiring the candidate comments, we classified them into five types  (\textit{design}, \textit{implementation}, \textit{defect}, \textit{test}, and \textit{documentation}) following da Silva Maldonado \textit{el at.} \cite{da2017using}. Initially, we intend to employ this information to guide annotators, thereby mitigating their workload. However, this information may have biased the annotators. 
Hence, we utilized this information to investigate TD types that are frequently misunderstood by the model's capabilities, thus pinpointing examples worthy of annotation (Section \ref{sec:sampling_annotation}). Instead of employing multiclass classification to detect TD types within comments, we constructed a binary classifier for each type. For instance, when considering TD types $X$, Classifier-$X$ is developed to distinguish whether a comment contains TD type $X$ or not. Similar to identifying comments containing TD, we designate training examples containing TD type $X$ as the positive class, while the remainder are categorized as the negative class. Consequently, 
we created five classifiers and each of the 1.6 million extracted comments was analyzed by these five classifiers to obtain pseudo-categories. Since these classifiers work independently, a comment can be categorized into more than one class. Figure \ref{fig:overlap_category} depicts the overlap categories predicted within a single comment, demonstrating the similarity between the two types of TD. It is shown that \textit{design} and \textit{implementation} are the two categories most often confused, a finding consistent with prior studies \cite{li2023automatic} that tried to merge these  categories. However, we maintained these categories separately for more fine-grained evaluation in our dataset.

\subsection{Sampling Strategy}
\label{sec:sampling_annotation}

The volume of detected comments (1.6M) is still very large, and resource-intensive for human 
relabeling. Therefore, we employed a technique to select useful samples for annotation.

Our objective is not only the selection of examples to explore the detection of TD at the function level but also the identification of comments that could enrich existing datasets. Following existing 
literature \citep{settles2009active, coleman2019selection, chitta2021training, sorscher2022beyond}, we utilized uncertainty scores to identify challenging instances for annotation, 
employing the Entropy score. Consider an instance $x$, a model with parameter $\theta$ and $C$ classes, the uncertainty score of the model on sample $x$ is 
as follows: 

\begin{equation} \label{eq:entropy}
\text{entropy}_\theta(x)=-\sum_{i=1}^C P_{\theta}(y=i|x) \log(P_{\theta}(y=i|x))
\end{equation}

This score indicates the confidence level of the trained model regarding a particular example: a low entropy score signifies high confidence in the model's prediction, whereas a high score indicates uncertainty. To enhance the existing datasets, we selected difficult examples for annotation. Combining the predictions from multiple TD classifiers, we devised a strategy to acquire a subset of functions for annotation, as outlined in Algorithm \ref{alg:sampling}. 
We constructed a candidate set comprising examples that imply more than one TD type using our five binary classifiers, along with examples exhibiting high uncertainty scores. We selected a subset of comments and corresponding functions from this set for annotation. After acquiring the set of functions, we extracted a list of comments $C_i$ corresponding to each function. These comments were subsequently relabeled and they serve as the primary information for defining TD types within the code functions.

\begin{algorithm}
\caption{Sampling informative subset for annotation}\label{alg:sampling}
\begin{algorithmic}
\Input
\Desc{$n$}{Number of samples for selection.}
\Desc{$c$}{List of TD types.}
\Desc{$N$}{No. comments extracted from detection tool.}
\Desc{$D$}{Extracted triplet \textit{(comment, function, category prediction list)} set: $\{d_i = (c_i, f_i, P_i) |$  $i=\overline{1,N}\}$.}
\Desc{$M$}{Dictionary of classifiers corresponding to each TD type key: $\{X$: Classifier-$X |$ $X \in \mathbcal{C}\}$.}
\EndInput\
\Output
\Desc{$\mathbcal{D}$}{Set of functions for annotation}
\EndOutput
\\
\State $Q \gets \{d_i|$ $d_i \in D, |P_i| > 1\}$  \Comment Comments that are predicted to contain more than 2 types of TD.
\State $\mathtt{UnSc} \gets \mathtt{list()}$
\For{$d_j$ in $D \backslash Q$}
    \State $p \gets P_j[0]$
    \State $\theta \gets \theta_{M[p]}$
    \State $s \gets \text{entropy}_\theta(c_j)$ \Comment Calculate using Equation \ref{eq:entropy}
    \State $UnSc$.add($s$)
\EndFor
\State $\mathtt{sID} \gets \mathtt{argsort}(\mathtt{UnSc}, \mathtt{desc=True})$
\State $\hat{Q} \gets \{d_j|$ $ d_j \in D \backslash Q, j \in {top}_{|Q|}(sID)\}$
\State $D \gets \mathtt{random\_sampling}_{n}(\{f_i |$ $ d_i \in Q \cup \hat Q\})$\\
\Return $D$
\end{algorithmic}
\end{algorithm}

\subsection{Data Annotation Process}
\label{sec:data_annotation}

\subsubsection{Annotation Group}

We formed an annotation team by hiring 7 final-year university students, each specializing in Software Engineering as their primary field of study. Their background allowed them to comprehend complex technical concepts and effectively apply this knowledge to the TD annotation process, thereby providing high-quality and contextually accurate data labels.

To further enhance their capabilities, we conducted a comprehensive training session tailored to the specific requirements of the labeling process. These sessions introduced the annotators to the field of TD and provided detailed explanations of various TD types, as specified in Section \ref{sec:Background}. By equipping the annotators with a thorough understanding of the task, we aimed to minimize errors and improve the overall quality of the labeling process.

\subsubsection{Labeling process}

Each annotator is provided detailed guidelines that serve as a reference throughout the annotation process. These guidelines include standardized procedures, examples of correctly labeled data, and common pitfalls to avoid.  By adhering to these guidelines, the annotators better maintain consistency and reliability across the dataset. Following da Silva Maldonado \etal \cite{maldonado2015detecting}, we developed a tool for the labeling process. 
However, diverging from the conventional approach of displaying only comments, we also included the corresponding code function as a reference for annotators. Annotators were asked to review both the comments and the corresponding code for labeling.  Moreover, we limited the labeling process to our five specific TD types: \textit{design}, \textit{defect}, \textit{documentation}, \textit{implementation}, and \textit{test} debts, in addition to \textit{non-SATD}. 
The students were given comment and corresponding code, and they had to read both and make a decision. In case, there is no TD contained in the code, so the corresponding label \textit{non-SATD} was given. 
Additionally, we conducted \textbf{Cross-checking }and \textbf{Label Auditing} to enhance the quality of the labeling process. Specifically, regarding a data sample, the labeling process is outlined as follows. 

\begin{enumerate}
    \item \textbf{Annotator assignment:} For each comment, two annotators were randomly chosen for labeling.
    \item \textbf{Cross-checking:} We collected the labels assigned to each example by the two annotators and made a comparison. If there was disagreement between the labeling results, we moved to the Label Auditing step; otherwise, the example was included in the final dataset.
    \item \textbf{Label Auditing:} We asked the two annotators to discuss their labeling, and reach an eventual consensus. 
\end{enumerate}

As shown in Table~\ref{tab:annotation_scores}, the labeling process was conducted in two phases. In the first phase, 1,400 comments were selected for annotation by seven annotators. Aiming for reliability, 
every comment was labelled by two students, \ie each annotator was assigned 400 comments. This phase 
helps the students familiarize themselves with the labeling task, and establish uniform conventions for the labeling process. Subsequently, in the second phase, there were 3,680 different comments, and each of them was independently evaluated by two students. This resulted in a total of 7,360 comments for the labeling process.  %
To guarantee the reliability of the labeling process, we assessed it using two consensus metrics: Raw Agreement and Inter-Annotator Agreement.

\begin{enumerate}
\item Raw Agreement: refers to the count of items for which both annotators assign identical labels, expressed as a percentage of the total items annotated \cite{artstein2017inter}.
\item Inter-Annotator Agreement (IAA):  Cohen's Kappa coefficient \citep{landis1977measurement, emam1999benchmarking} was applied to quantify the agreement or consistency between different annotators.
\end{enumerate}

Table \ref{tab:annotation_scores} presents the annotation scores across two phases. In the initial phase, as the annotators were becoming acquainted with the task, there is a relatively low agreement of 37\%, referring to a \textit{Fair} agreement. However, following reviews and discussions, the agreement strength improved significantly. In the second phase, the Raw Agreement increased by over 35\%, and the IAA improved by over 8\%, resulting in a \textit{Moderate} agreement strength. This demonstrates the reliability of our labeling process and the overall quality of the dataset. Following the labeling of TD types for the comments, we aligned the labeled comments with their corresponding code functions, thus obtaining multi-TD type information for an entire code function.

\section{Data Characteristics}
\label{sec:Datasets}


\begin{table}[t]
\centering
\caption{The comparison between popular SATD benchmarks and \TS. NL-sample refers to data in natural language text format, such as comments, pull requests, issues, and commit messages.}
\begin{tabular}{l|c|c|c|c}
\toprule
Dataset & \#Code-sample & \#NL-sample & \% TD NL-samples & \#Repo \\
\midrule
\Maldonado \cite{da2017using} & - & 62,566 & 6.5 & 10 \\
4Source-SATD \cite{li2023automatic} & - & 95,455 & 8.5 & 103 \\
SATD in R \cite{sharma2022self} & - & 146,583 & 3.4 & 503 \\
\midrule
\TS & \textbf{1,255} & 4,981 & \textbf{31.1} & \textbf{974} \\
\bottomrule
\end{tabular}
\label{tab:data_statistic}
\end{table}

To support detecting technical debt in both comments and code, we constructed two datasets.

\subsection{Dataset for TD detection in Source Code}
\label{sec:tscode}

We introduce a dataset named \TScode,  to support detecting technical debt in  source code  without relying on natural language comments. Unlike comments, a function can contain multiple types of technical debt; hence, we formulate this scenario as a multi-label classification problem. Specifically, we exclude comments within the function and consider TD types that indicate intrinsic issues in the source code: \textit{design}, \textit{implementation}, \textit{defect}, and \textit{test}. As a result, \TScode presents a challenge for detecting these types of TD within a code function.  Table \ref{tab:data_statistic} highlights that no existing dataset has addressed this crucial scenario, underscoring the significance of \TScode. 

\begin{figure}[t!]
\centering
\small
\includegraphics[width=0.68\textwidth]{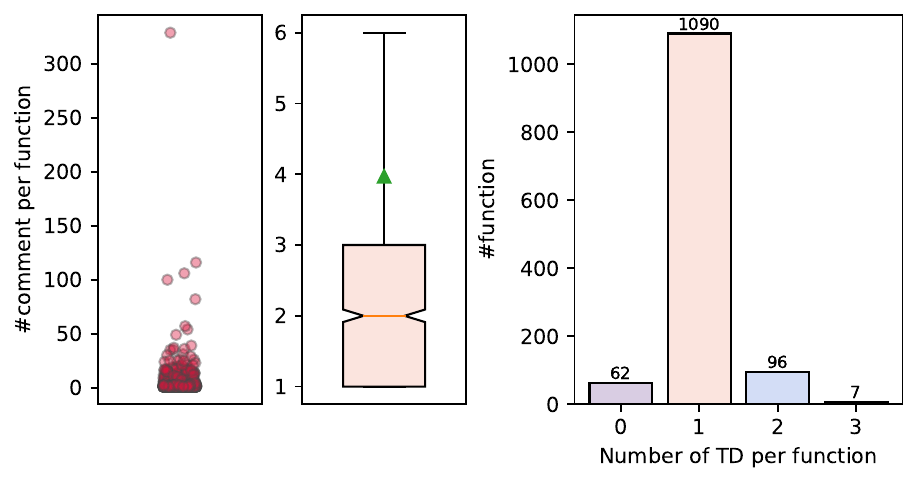}
\caption{Statistics of \TScode. Left: Distribution of the number of comments per function. Right: Distribution of the number of TD types within a function.}
\label{fig:code_distribution}
\end{figure}

\TScode includes 1,255 Java functions from 974 projects. Figure \ref{fig:code_distribution} indicates that the average function contains two comments. More than 1,000 functions (86.9\%) contain only a single type of TD. 
Only 7 functions contain three types of TD, and none encompass all types. This suggests that TD is mostly homogeneous within a function. %
Furthermore, although \textit{non-SATD} comments are the majority, only a small portion of the functions exhibit no type of TD. This is because we specifically focused on selecting functions with TD comments to facilitate more efficient labeling, as outlined in Section \ref{sec:sampling_annotation}.

\subsection{Dataset for SATD-Related Tasks}
\label{sec:tscomment}

Since we construct \TScode using information from comments and later exclude these comments, preserving them can help support existing SATD datasets. Therefore, we created \TScomment, where comments serve as the input source, to support SATD-related tasks, including the identification, classification, and detection of TD. These tasks are structured as multi-class classification problems, with details in Section \ref{sec:RQ1}. In addition, unlike existing datasets (Table \ref{tab:data_statistic}), each comment in \TScomment is associated with its corresponding code, providing a richer context for investigation and analysis. 

\begin{figure}[t]
\centering
\small
\includegraphics[width=0.8\textwidth]{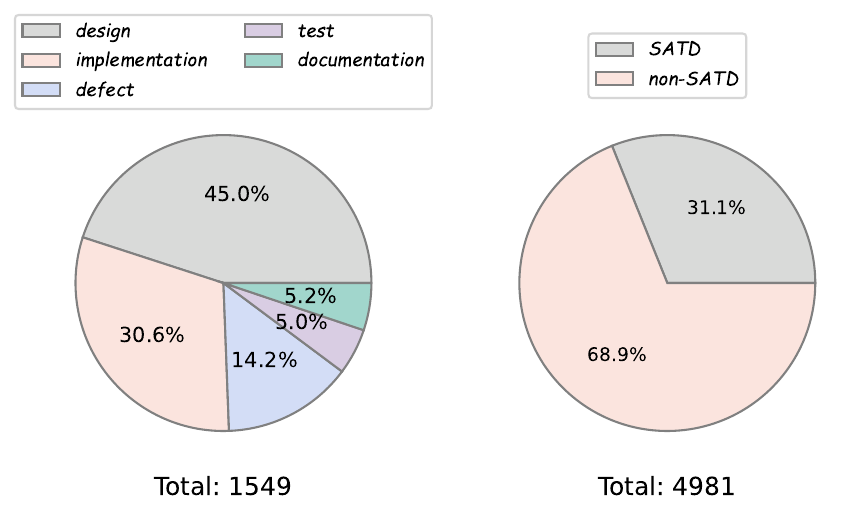}
\caption{Category distribution in \TScomment. Left: distribution of TD categories within comments containing SATD. Right: percentage of comments that contain versus those that do not contain SATD.}
\label{fig:data_distribution}
\end{figure}

Figure \ref{fig:data_distribution} presents the statistics of the \TScomment dataset, which contains 5,000 labeled comments across six categories: the five TD types and \textit{non-SATD}.
Consistent with previous studies \cite{maldonado2015detecting, da2017using, li2023automatic}, \textit{design} and \textit{implementation} debts constitute the majority, with fewer entries for \textit{test} and \textit{documentation}, reflecting the real-world distribution. On the other hand, comments with SATD represent a significant portion of the dataset, \ie 31.1\%, which is considerably higher than the proportions in previous datasets (Table \ref{tab:data_statistic}). This underscores the effectiveness of our SATD detection tool in identifying SATD comments, 
which helps to mitigate the imbalance between SATD and \textit{non-SATD} comments. 
Table \ref{tab:data_statistic} shows that our dataset is sourced from 974 repositories. Compared to existing studies  \TScomment is derived from a more diverse range of sources,  capturing a wider variety of commenting and coding styles.

\section{Experimental Results}
\label{sec:Experiment}



\revised{We now discuss the experimental results and answer
the research questions outlined in Section~\ref{sec:introduction}. 
\textbf{RQ$_1$} seeks to assess the significance of source code in detecting SATD comments. We use the function containing the comments as supplementary features to aid in identifying SATD comments.
Lastly, we assess the effectiveness of various models in identifying technical debts from source code \textit{without} relying on SATD comments. In \textbf{RQ$_2$} our goal is to utilize the curated dataset to detect the presence of different types of technical debts within source code. For \textbf{RQ$_3$} we demonstrate that the enhanced dataset improves the performance of existing SATD classification models.}

\subsection{\rqone}
\label{sec:RQ1}

\noindent \textbf{Motivation:} As mentioned earlier, most existing work on TD detection has primarily focused on comments and other textual artifacts such as commits and issues~\citep{maldonado2015detecting,da2017using, li2023automatic, 9226330, sala2021debthunter} while overlooking the content of source code.
However, certain features of source code can indicate the presence of some types of technical debt, such as \texttt{design} or \texttt{implementation}.
For instance, \texttt{design} technical debt may be introduced in anti-pattern source code where developers have neglected specific design principles \cite{xiao:tse2022}. This highlights the potential of source code to provide meaningful features for the detection of technical debt. Furthermore, during the annotation process, it was observed that annotators were more proficient in identifying technical debt when utilizing source code information as a reference, rather than relying exclusively on comments. \revised{A few recent studies have tried to leverage source code alongside comments to improve SATD detection, demonstrating that code context can enhance model performance \citep{10.1145/3524842.3528469, zampetti2020automatically, russo2025leveraging}. However, these approaches often lack a comprehensive analysis into critical factors such as the effect of different code-comment integration techniques or the optimal scope of code context--instead frequently relying on simplistic strategies of using full code. Therefore, this RQ aims to evaluate the effectiveness of incorporating source code into SATD detection and to systematically investigate various factors to identify the most effective strategies for utilizing code context.}

\vspace{.2cm}
\noindent \textbf{Approach:} 
We assess the effectiveness of different models in detecting SATD comments by comparing their performance when using only comments with that 
when incorporating both comments and source code. We used \TScomment for these experiments. To ensure a comprehensive evaluation, we designed the experiment based on the following considerations.

\begin{itemize}[leftmargin=*]
    \item \textit{Model versatility:} 
    Our goal is to explore the impact of incorporating source code 
    in various models to demonstrate the effectiveness of this approach. As such, we apply the proposed method to four PLMs:  RoBERTa, CodeBERT, UniXCoder and GraphCodeBERT. BERT has been excluded because it showed performance similar to or lower than its variant, RoBERTa as can be seen in Table~\ref{tab:rq1}.

    \vspace{.2cm}
    \item \textit{Integration techniques to combine source code and comments:} 
    We explore the effectiveness of two distinct methods for combining source code and comments. Specifically, both the source code and comments are tokenized into separate sequences of tokens, which are then integrated using two different strategies. 
 
    \begin{enumerate}[label=(\alph*)]
        \item String Concatenation (StrConcat): 
        Two sequences of tokens are concatenated then passed into pre-trained models.
        This approach is typically employed in prior studies for text classification tasks~\citep{devlin2018bert, liu2019roberta, lan2019albert}.
        \item Code Attention (CodeAtt): 
        The source code and comments are processed independently using a pre-trained encoder, resulting in an embedding vector for each token. After that, we calculate the attention score for each code token in relation to the embedding representation of the comment tokens.
        Let $G_{M \times D}$ represent the embedding matrix for $M$ code tokens from the source code, and $H_{N \times D}$ represent the embedding matrix for $N$ comment tokens, where $M, N$ are the number of code and comment tokens, respectively, $D$ is the size of the embedding vector. The final embedding that combines both source code and comment is obtained by taking the dot product of the code's attention matrix with the comment embedding, as shown below.
        \begin{gather}
            A = softmax(G \cdot H^T) \nonumber \\
            classification\_emb = A \cdot H^T \nonumber
        \end{gather}
    \end{enumerate}

    \vspace{.2cm}
    \item \textit{Code context scope:} 
    Based on our observations during the labeling process, annotators did not need to review the entire code to determine the type of technical debt (TD) associated with a comment; they only needed to scan the nearby code. With this in mind, we investigated the impact of varying the length of the code context. Specifically, we assessed the effect of using the surrounding code by including 2, 10, and 20 lines, as well as the entire function. Figure \ref{fig:extraction} illustrates an example of utilizing a code context that includes 2 lines of code.
\end{itemize}

\begin{table}[t!]
\centering
\caption{Performance (F1-score) comparison of various models on SATD detection using only comments versus incorporating additional source code. The subscripts 
accompanying the numerical results indicate the number of context lines that produced the best outcomes for each model, with \textbf{ff} representing the use of the full function.}

\begin{tabular}{l|cccc}
\toprule
\multirow{2}{*}{Model} & Comment & \multicolumn{3}{c}{Comment + Code} \\ 
\cline{3-5}
& only & StrConcat & CodeAtt & Ensemble \\
\midrule
RoBERTa & 68.28	& $69.11_2$ ({\color{darkgreen} $\uparrow$1.22\%}) & $69.45_2$ ({\color{darkgreen} $\uparrow$1.71\%}) & \textbf{69.92}\\
CodeBERT & 72.75 & $76.30_{\text{ff}}$ ({\color{darkgreen} $\uparrow$4.88\%}) & $74.80_{\text{ff}}$ ({\color{darkgreen} $\uparrow$2.82\%}) & \textbf{76.55} \\
UniXCoder & 70.62 & $71.54_{20}$ ({\color{darkgreen} $\uparrow$1.30\%}) & $71.46_{\text{ff}}$ ({\color{darkgreen} $\uparrow$1.19\%}) & \textbf{72.22}\\ 
GraphCodeBERT & 71.39 & $74.86_2$ ({\color{darkgreen} $\uparrow$4.86\%}) & $72.75_{\text{ff}}$ ({\color{darkgreen} $\uparrow$1.91\%}) & \textbf{75.25}\\ 
\bottomrule
\end{tabular}
\label{tab:rq2}
\end{table}

\vspace{.2cm}
\noindent \textbf{Result:} Table \ref{tab:rq2} presents the performance of the four PLMs in detecting SATD comments, comparing the outcomes of using only comments versus combining comments with code context. The analysis of the results is based on the three previously mentioned aspects.

\begin{itemize}[leftmargin=*]
    \item \textit{Model versatility:} The results shows that incorporating code context significantly improves the performance across all evaluated models and integration approaches. Notably, CodeBERT achieves the highest accuracy 
    when using comments alone and exhibits the greatest improvement, attaining an F1-score of 76.3\% with the addition of code context. These findings highlight the value of integrating source code information for detecting SATD, boosting the effectiveness of already high-performing models. Furthermore, the improvements observed across different models further illustrate the robustness and the adaptability of this approach in identifying SATD.
    
    \vspace{.2cm}
    \item \textit{Integration techniques:} Overall, both proposed methods improve performance across all models, with StrConcat demonstrating greater effectiveness than CodeAtt except in the case of RoBERTa. Specifically, StrConcat enhances the performance of CodeBERT and GraphCodeBERT by over 4.88\% and 4.86\%, respectively. In contrast, the improvements achieved with CodeAtt are more moderate, ranging between 1.19\% and 2.82\%. 
    The superior results of StrConcat highlight the ability of Transformer-based models to effectively process multi-modal inputs through their self-attention mechanism, a capability that has been demonstrated in various downstream tasks~\cite{wei2020retrieve,wang2022bridging,zhou2023qtc4so}. 

    \vspace{.2cm}
    \item \textit{Code context scope:} Table \ref{tab:rq2} shows that the F1-Score of four models is significantly improved when using either two lines of code context or the entire function code. 
    The tendency to favor two surrounding code lines during prediction aligns with human intuition, which relies on local context for annotation.
    Moreover, incorporating the entire function code highlights the capability of these models to leverage global context.
    This context could enhance the performance by enabling models to identify relevant code snippets across the function, offering a more nuanced understanding of the function’s structure and semantic, thereby potentially improving the reliability of the models. 
    While employing code context generally demonstrates improvements over using comments alone, varying the context length might impact model performance differently. 
    Consequently, we employ an ensemble approach to combine model predictions across different code context lengths. Specifically, for each model 
    a majority voting mechanism was applied to produce the final prediction. Each model was configured with varying code context lengths, including 2, 10, 20 lines and the entire function code, considering  CodeAtt as the input concatenation approach for RoBERTa and StrConcat for the other models. As shown in the last column of Table~\ref{tab:rq2}, this ensemble approach achieves the highest performance across all four models, underscoring the advantage of leveraging multiple code context lengths for identifying SATD comments.
\end{itemize}


\begin{tcolorbox}[colback=gray!5!white,colframe=darkgray,title=Answer to RQ1]
\begin{itemize}[leftmargin=*]
    \item Incorporating comments with source code information results in performance improvements across various models 
    compared to using comments alone, highlighting the robustness, versatility, and adaptability of this approach in detecting SATD.
    
    \item The proposed methods, StrConcat and CodeAtt, effectively utilize the source code context and enhance the performance across all evaluated models. This paves the way for future research 
    with the ultimate aim of further improving the prediction. 
    
    \item When comment and code context are combined as input, 
    an optimal performance is achieved with 
    2 surrounding code lines or by including the entire code function. Combining various scopes demonstrates the effective contribution of each scope, highlighting the potential of multi-code scope strategies in improving SATD detection. 
    
\end{itemize}
\end{tcolorbox}

\subsection{\rqtwo}
\label{sec:RQ2}

\textbf{Motivation:} As mentioned earlier, technical debt is frequently identified through textual content, such as comments or issue reports. However, when such debt is not explicitly documented, existing tools are unable to detect it, despite its presence in the code. Moreover, many comments become outdated or inconsistent with the actual code, as developers often fail to update comments after modifying the code. This discrepancy between comments and code introduces a significant blind spot for tools that rely solely on textual indicators, limiting their ability to accurately detect technical debt. \revised{Besides, some prior studies have explored technical debt detection directly within source code; however, they typically focused on limited classification schemes--such as distinguishing only between high and not high TD \citep{10586898, 9622154}, or on identifying the mere presence or absence of code smells \citep{yadav2024machine}, without capturing the broader diversity of technical debt types}. Therefore, there is a pressing need for more advanced approaches that surpass textual cues to effectively identify and manage technical debt within code bases. In response, we propose investigation on detecting multi-type TD in source code, extending beyond conventional binary classification frameworks.




\vspace{.2cm}
\noindent \textbf{Approach:} We designed a scenario where TD detection relies solely on source code. 
Specifically, we constrained the scope to the function level for practical application, as analyzing the entire file is lengthy and challenging for users to segment after detecting TD. We utilized \TScode for our experiments, addressing it as a multi-label classification problem. In order to extract information from source code, we have investigated different PLMs, categorized into three architectures as detailed in Section \ref{sec:plms_type}.

\begin{itemize}
    \item Encoder-based PLMs: Language models that are based on Transformer architecture utilizing the Encoder layer. 
    \item Encoder-Decoder-based PLMs: Language models, built on top of Transformer architecture, that leverage both the Encoder and Decoder layers.
    \item Decoder-based PLMs: Language models that use Decoder layer of Transformer architecture, trained with Causal Language Modeling. The models, characterized by a large number of parameters (in billions), are commonly referred as Large Language Models (LLMs).
\end{itemize}

\begin{table}[t!]
\centering
\caption{Performance of different PLMs on TD detection using \TScode.}
\begin{tabular}{l|c|cc}
\toprule
Model & Model size & EM & F1 \\
\midrule
\multicolumn{4}{c}{Encoder-based PLMs} \\
\midrule
CodeBERT \cite{DBLP:conf/emnlp/FengGTDFGS0LJZ20} & 125M  & 38.28	& 43.47  \\
UniXCoder \cite{DBLP:conf/acl/GuoLDW0022} & 125M 
 & 38.12 & 42.58	 \\
GraphCodeBERT \cite{guo2020graphcodebert} & 125M  & \underline{39.38} & \underline{44.21} \\
RoBERTa \cite{liu2019roberta} & 125M  & 35.37	&  38.22	 \\
ALBERT \cite{lan2019albert} & 11.8M & 39.32 & 41.99 \\
\midrule
\multicolumn{4}{c}{Encoder-Decoder-based PLMs} \\
\midrule
PLBART \cite{ahmad2021unified} & 140M & 36.85 &	 39.90 \\
Codet5 \cite{wang2021codet5} & 220M & 32.66 & 35.41	 \\
CodeT5+ \cite{wang2023codet5+} & 220M & 37.91	&  41.96	 \\
\midrule
\multicolumn{4}{c}{Decoder-based PLMs (LLMs)} \\
\midrule
TinyLlama \cite{zhang2024tinyllama}  & 1.03B & 37.05 & 40.05 \\
DeepSeek-Coder \cite{zheng2024opencodeinterpreter}  & 1.28B & \textbf{42.52} & \textbf{46.19} \\
OpenCodeInterpreter \cite{guo2024deepseek}  & 1.35B & 38.16 & 41.76 \\
phi-2 \cite{roziere2023code} & 2.78B & 37.92 & 41.57	\\
starcoder2 \cite{lozhkov2024starcoder} & 3.03B & 35.37 & 41.77	\\
CodeLlama \cite{roziere2023code} & 6.74B & 34.14 & 38.16	 \\
Magicoder \cite{wei2023magicoder} & 6.74B & 39.14 &	42.49\\
\bottomrule
\end{tabular}   
\label{tab:rq3}
\end{table}

We conducted experiments on 16 models, including 5 Encoder-based PLMs, 3 Encoder-Decoder-based PLMs, and 8 Decoder-based PLMs. 
Since only code is used 
as input, and Sections~\ref{sec:RQ1} and~\ref{sec:RQ2} demonstrated the superior performance of code-based PLMs, we experiment with models primarily pre-trained on coding corpora.
The language model head layer is replaced with a linear classification head during fine-tuning on the downstream task. All models are fine-tuned for 10 epochs using a batch size of 32 and a learning rate of $1e-5$. For models with more than 6 billion parameters, we utilize LoRA \cite{hu2021lora} with a learning rate of $1e-3$ for fine-tuning due to resource constraints. For decoder-base PLMs, we apply a template for input presented in the online appendix,  
and use the embedding of final tokens as the representation fed into the classification head. We randomly split \TS into 10 folds for cross-validation and report the average Exact Match (EM) and F1-score across these folds for each model. 

\begin{figure}[h!]
    \centering
    \includegraphics[width=0.8\textwidth]{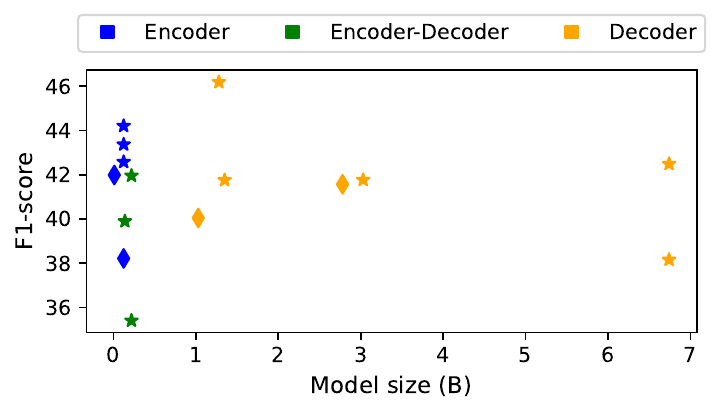}
    \caption{F1-score of various PLMs on \TScode across different model sizes, types, and pretraining datasets. $\blacklozenge$ denotes NL-based PLMs; $\filledstar$ represents code-based PLMs.}
    \label{fig:rq3}
\end{figure}

\noindent \textbf{Result:} Table \ref{tab:rq3} shows the experimental results, with  \textbf{bold text} indicating the highest score, while \underline{underlined scores} representing the runner-up. DeepSeek-Coder achieves the best performance with an F1-score of 46.19\% marking an improvement of 4.98\% over the second highest one, \ie 
GraphCodeBERT getting 44.21\%. Though previous studies indicated a limited adaptability of LLMs to classification tasks \citep{zhang2024pushing, sun-etal-2023-text}, these results highlight the potential of such models. However, models containing the Decoder module generally exhibit lower performance compared to Encoder-based models. Figure \ref{fig:rq3} supports this observation, as the three models following DeepSeek-Coder are all Encoder-based. Several factors can account for this observation. Firstly, Encoder-Decoder and Decoder-based models are 
pretrained on generation tasks, which may result in suboptimal performance on classification tasks due to the lack of task-specific optimization. Secondly, some studies \citep{behnamghader2024llm2vec, lee2024nv} showed that Decoder-based models are less effective for text representation or embeddings due to their causal attention mechanism, which limits the model’s ability to learn robust representations. Hence, the embedding information before the classification head is not sufficiently rich, leading to a suboptimal performance. Though Encoder-based models experience a slight performance drop compared to that of DeepSeek-Coder, they achieve this with significantly fewer parameters--around 90\% less--offering a more practical approach to TD detection using source code.

Figure \ref{fig:rq3} further 
highlights the superior performance of code-based PLMs compared to  NL-based PLMs when considering models of comparable size. For example, within the group of models containing 100M to 200M parameters, GraphCodeBERT achieves the highest F1-score of 44.21\%. Similarly, models with sizes around 1B and 3B parameters also exhibit the best performance with two code-based PLMs, DeepSeek-Coder and StarCoder2, respectively. This further reinforces the superiority of code-based PLMs in this scenario. However, the performance of all models remains below 50\% in both EM and F1-score, indicating the need for more advanced approaches and further improvements.

\begin{tcolorbox}[colback=gray!5!white,colframe=darkgray,title=Answer to RQ2]
\begin{itemize}[leftmargin=*]
    \item DeepSeek-Coder 
    achieves the highest accuracy on \TScode. 
    In contrast, Encoder-based models exhibit a slight performance drop 
    but with a substantial reduction in parameters, making them more practical in real-world scenarios.  
    
    \item Code-based PLMs show superiority in the detection of TD from source code. However, their performance remains below 50\% in both EM and F1-score. This indeed highlights the need for more advanced methods and further research.
\end{itemize}
\end{tcolorbox}

\subsection{\rqthree}
\label{sec:RQ3}

\begin{table*}[t!]
\centering
\caption{The performance (F1-score) of five PLMs across three tasks when trained on the M-62K (\Maldonado) dataset and further enhanced with the additional \TS dataset.}
\begin{tabular}{l|c|c|c|l|c|l}
\toprule
\multirow{2}{*}{Model} & \multicolumn{2}{c|}{Identification} &  \multicolumn{2}{c|}{Classification} &  \multicolumn{2}{c}{Detection}\\
\cmidrule{2-7}& M-62K & +\TS ($\Delta$) & M-62K & +\TS ($\Delta$) & M-62K & +\TS ($\Delta$) \\
\midrule
BERT & 87.96 & {\bf88.99} ({\color{darkgreen} $\uparrow$1.17\%}) & 51.42 & {\bf55.10} ({\color{darkgreen} $\uparrow$7.16\%}) & 45.99 & {\bf49.64}  ({\color{darkgreen} $\uparrow$7.94\%}) \\
RoBERTa & 89.06 & {\bf89.96} ({\color{darkgreen} $\uparrow$1.01\%}) & 53.91 & {\bf55.32} ({\color{darkgreen} $\uparrow$2.62\%}) & 46.13 & {\bf52.86} ({\color{darkgreen} $\uparrow$14.59\%}) \\
UniXCoder & 88.38 & {\bf88.42} ({\color{darkgreen}$\uparrow$0.05\%}) & 54.82 & {\bf54.99}  ({\color{darkgreen} $\uparrow$0.31\%}) & 50.94 & {\bf52.11} ({\color{darkgreen} $\uparrow$2.30\%}) \\
CodeBERT & 88.74 & \textbf{90.06} ({\color{darkgreen} $\uparrow$1.49\%}) & 57.50 & \textbf{63.70} ({\color{darkgreen} $\uparrow$10.78\%}) & 53.79 & \textbf{55.60} ({\color{darkgreen} $\uparrow$3.36\%})\\
GraphCodeBERT & 89.94 & \textbf{90.12} ({\color{darkgreen} $\uparrow$0.20\%}) & 58.00 & \textbf{60.44} ({\color{darkgreen} $\uparrow$4.21\%}) & 49.12 & \textbf{56.87} ({\color{darkgreen} $\uparrow$15.78\%})\\

\bottomrule
\end{tabular}
\label{tab:rq1}
\end{table*}

\noindent \textbf{Motivation:} 
Our primary focus in this RQ is to assess whether the newly identified SATD comments from the Stack corpus contribute positively to the performance of SATD comment detection. Our objective is to enhance the state-of-the-art benchmark dataset  introduced by da Silva Maldonado \etal~\cite{da2017using}, henceforth referred to as \Maldonado. This dataset comprises 62,566 comments, of which 4,071 (approximately 6.5\%) encompass one of five categories of technical debt. While the dataset's volume is relatively limited, making it challenging to apply deep learning models, it is crucial to increase the diversity of data samples within each category to enhance the training performance of SATD detection models.

\vspace{.2cm}
\noindent \textbf{Approach:} 
For this RQ, our objective is to examine whether manually classified comments from the \TScomment dataset improve the detection of SATD comments. Following previous studies~\cite{da2017using,ren2019neural,sala2021debthunter,sheikhaei2024empirical}, we employed a cross-project experimental approach. The benchmark dataset \Maldonado comprises 10 projects, divided into 10 validation folds. We used one fold (one project) for testing and the remaining nine folds (nine other projects) for training.
To prevent data leakage, we removed duplicate entries from the entire dataset before splitting it, resulting in 38,269 samples. We then analyzed the impact of incorporating \TS during training, denoted as $+$\TS, compared to using only the state-of-the-art dataset, \Maldonado, across three scenarios: SATD-Identification, SATD-Classification and SATD-Detection.

\begin{enumerate}
\item SATD Identification ($\mathbb{S}_1$): This task involves determining the presence 
of technical debt in a given comment.
\item SATD Classification ($\mathbb{S}_2$): In this task we categorize comments identified as containing SATD into one of five distinct categories.
\item SATD Detection ($\mathbb{S}_3$): This task combines both identification and classification, categorizing each comment into one of six groups: five for the types of technical debt and one for \textit{non-SATD}.
\end{enumerate}

For each scenario, we used the same test set while training the model separately with the \Maldonado and with $+$\TS datasets. Research in the field of SATD detection has highlighted the promising results of PLMs~\cite{9804499,sharma2022self,sheikhaei2024empirical}, thus we also employed these models to detect SATD comments. Specifically, we experimented with
BERT~\cite{devlin2018bert}, RoBERTa~\cite{liu2019roberta}, CodeBERT~\cite{DBLP:conf/emnlp/FengGTDFGS0LJZ20}, UniXCoder~\cite{DBLP:conf/acl/GuoLDW0022} and GraphCodeBERT~\cite{guo2020graphcodebert}. PLMs are employed as comment encoders, followed by a fully connected network dedicated to downstream tasks such as identification, classification, and detection, as previously described. All models are fine-tuned for 10 epochs using a batch size of 32 and a learning rate of $1e-5$. Given the significant class imbalance in the dataset, the F1 score to was used to evaluate performance.

\vspace{.2cm}
\noindent \textbf{Result:} Table \ref{tab:rq1} depicts the performance of five different PLMs across three aforementioned scenarios $\mathbb{S}_1$, $\mathbb{S}_2$ and $\mathbb{S}_3$. 
We see that incorporating \TS during the training phase consistently boosts the performance of all PLMs across all 
tasks. 
Specifically, the identification of SATD comments from the entire set of comments demonstrates an improvement ranging from 0.05\% to 1.49\% across all models when the training phase is supplemented with the \TS dataset, as opposed to relying solely on the \Maldonado dataset. 
In this scenario, the dataset demonstrates a considerable imbalance, with \textit{non-SATD} data making up more than 90\% of the total. The improvement highlights the effectiveness of the \TS$_{comment}$ in mitigating the data imbalance issue, and this is significant because real-world scenarios will have a similar imbalance. In the context of SATD detection, BERT and RoBERTa exhibited lower performance compared to the other three code-based PLMs. 
However, training with $+$\TS considerably improves the performance of these two models, leading to comparable results across all the models.
For instance, applying $+$\TS during training improved BERT's performance by 7.94\% and RoBERTa's by 14.59\% in detecting various types of SATD comments. A similar trend is observed in other 
scenarios, where all models show enhanced performance when utilizing $+$\TS.
Specifically, the performance in classifying the five SATD types increases by approximately 0.31\% to 10.78\% for all PLM models.
Additionally, the results indicate that CodeBERT and its variant, GraphCodeBERT, achieve the highest performance across all scenarios, highlighting the advantage of code-based PLMs for the detection of SATD comments.

\begin{figure}[t!]
\centering
\small
\includegraphics[width=0.92\textwidth]{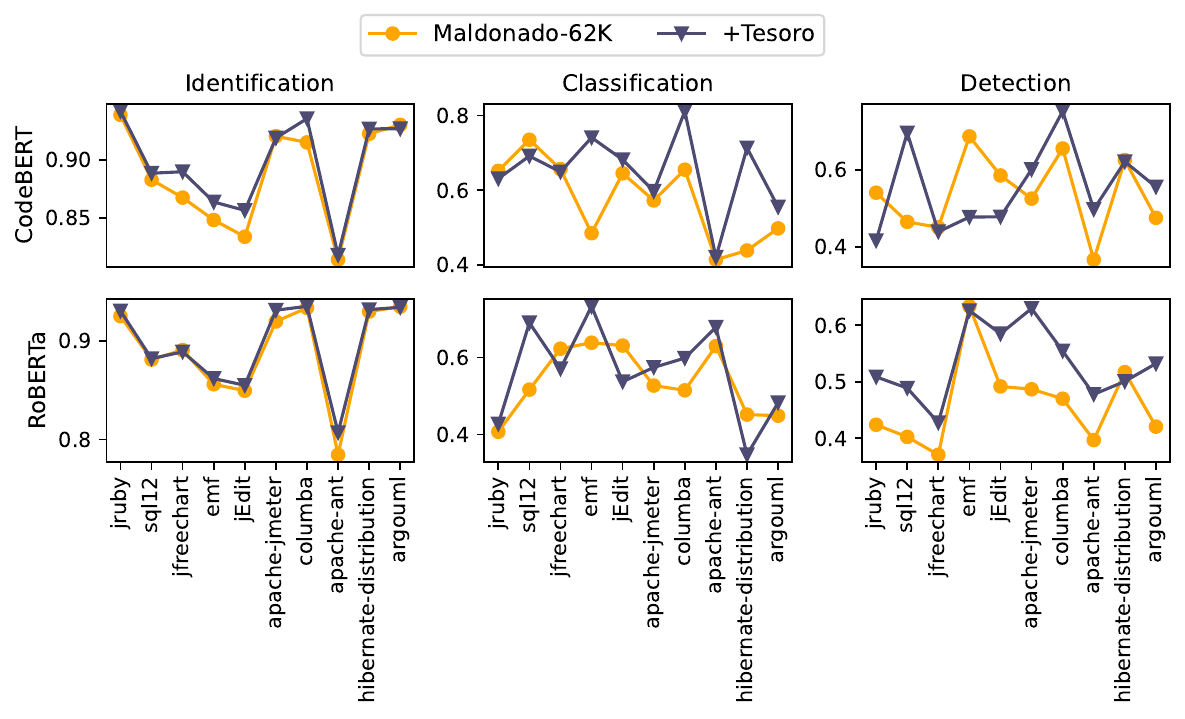}
\caption{An in-depth analysis of CodeBERT and RoBERTa performance across three scenarios for 10 projects.}
\label{fig:rq1_projects}
\end{figure}

Finally, we conducted an in-depth analysis of the results from 10 large-scale open source projects across all three scenarios. In this setup, each project serves as a test fold, while the PLMs are trained on the remaining 9 projects.
For illustrative purposes, we present the findings for CodeBERT and RoBERTa, as depicted in Figure~\ref{fig:rq1_projects}.
Overall, it is evident that training the models with $+$\TS results in improved performance across all projects.
In particular, in the $\mathbb{S}_1$ task, the performance improved across all tested projects for both models, highlighting the enhanced training set's effectiveness in addressing the imbalance issue in the original dataset.
Furthermore, 7 out of the 10 folds show an increase in F1-score when training CodeBERT on classification and detection tasks with $+$\TS, while RoBERTa exhibits significant improvement in 9 out of 10 test sets for the SATD detection task.

\begin{tcolorbox}[colback=gray!5!white,colframe=darkgray,title=Answer to RQ$_3$]
\begin{itemize}[leftmargin=*]
    \item 
    There is 
    an 
    improvement in the prediction performance when \TS is incorporated into the training, validating the efficacy of our data pipeline in selecting informative samples, and proving a robust annotation process.
    \item CodeBERT and GraphCodeBERT consistently achieve superior performance and notable improvements across all three tasks, highlighting the advantages of employing code-based PLMs for SATD comment detection.
\end{itemize}

\end{tcolorbox}




\section{Discussion}
\label{sec:Discussion}

We now discuss possible impacts of our work, and highlight the threats to validity of the findings.

\subsection{Implications}

Our work has the following implications:

\begin{itemize}
    \item Unlike existing approaches, which rely solely on textual data such as issue reports, comments, commits to detect TD, we propose using source code as a means to facilitate such the detection.  
    This may have significant implications in practice, as source code and the associated text might not be coherent, and using only source code helps us capture the intrinsic debt, without relying on the presence of any 
    accompanying textual data.  
    With the curated datasets, we expect to lay the foundations for a new method to detect TD. 
    This may be beneficial to industry, as software companies could make use of our datasets to train tailored machine learning models to recognize TD directly from their source code. This would help them save time and effort, thereby increasing the overall productivity. 
    
    \item The results in \textbf{RQ$_2$} show that the PLMs we considered achieve mediocre results when detecting TD in source code. This implies that there is still room for improvement: more advanced detection models are needed. We anticipate that 
    LLMs could be an eventual solution to this problem, as they have been trained with a huge amount of data including source code, with the potential to better capture the intrinsic features of TD contained in source code. This, however, needs further refinement and empirical evidence and is part of our future work.
    
    \item The curated dataset is expected to advance research in technical debt detection from source code, and holds the potential to facilitate the identification of other software artifacts, such as code smells. 
    
\end{itemize}

\subsection{Threats to validity}
We see the following threats to validity related to this research:

\begin{itemize}
    \item \textbf{Internal validity.} This threat is related to the confounding factors that might impact the validity of the evaluation results. The dataset that we used to train the classifier to look for additional SATD comments could cause the engine to harvest false positives if it is not properly curated. To mitigate this threat, we used the preexisting \Maldonado dataset, which was carefully classified, and has been utilized in various studies. When conducting the user study, we tried to avoid any bias by involving seven Computer Science students with significant programming experience in the manual evaluation step. In addition, the results of each student were then double checked by another student to resolve any conflicts and to increase the reliability of the results.
    
    \item \textbf{External validity.} This threat concerns the generalizability of our findings. We used SATD comments extracted from projects to train a classifier to locate Java source code containing TD. While it seems reasonable to assume that this method could also be used to identify TD contained in other programming languages, we have not shown that. In this paper we managed to validate our hypothesis only on code written in Java. Therefore future work is required to further validate the generalizability of our pipeline to other languages.

    
\end{itemize}

\section{Related Work}
\label{sec:RelatedWork}


This section reviews related studies that address the problem of technical debt (TD). It also highlights various datasets, approaches, and methods developed to identify and detect TD across different software systems.

\vspace{0.2cm}
\noindent \textbf{Technical debt detection datasets:} The availability of datasets is critical for advancing research in technical debt related tasks, providing empirical evidence to validate the effectiveness of detection techniques. Large-scale studies of TD frequently focus on a specific category, with Self-Admitted Technical Debt (SATD) being a prevalent area of focus. Maldonado \textit{et al.} \cite{maldonado2015detecting, da2017using} presented a widely utilized dataset, comprising more than 62,000 comments from 10 Java projects, classified into five types of TD and \textit{non-SATD}. Instead of using only comment, Li \textit{et al.} \cite{li2023automatic} introduced the dataset by investigating four different sources: code comments, commit messages, pull requests, and issue tracking systems. In addition, they merged \textit{ code debt} and \textit{design debt} into a single category due to their high similarity. To explore programming languages beyond Java, Sharma \textit{et al.} \cite{sharma2022self} introduced a dataset for examining SATD in the R language, which includes over 140,000 samples and expands the number of TD types to 12. Furthermore, several datasets \cite{palomba2015landfill, madeyski2020mlcq, nandani2023dacos, kovavcevic2024automatic} have been introduced to address code smells, which is a related issue to TD. 

\vspace{0.2cm}
\noindent \textbf{SATD detection techniques:} Early methods for identifying and detecting SATD relied on rule-based approaches that searched for matching keywords or phrases \cite{guo2021far, sridharan2023pentacet}. For example, the Matches Task Annotation Tags (MAT) \cite{guo2021far} method demonstrated that simply matching a set of commonly used task annotation tags, such as \textit{TODO}, \textit{FIXME}, \textit{HACK}, and \textit{XXX}, can achieve a significant performance in identifying SATD. In PENTACET, Sridharan \textit{et al.} \cite{sridharan2023pentacet} built upon the 64 SATD detection patterns initially introduced by Potdar and Shihab \cite{potdar2014exploratory}. By leveraging the Sense2Vec tool, they expanded the set to 1,041 patterns, significantly increasing the scope for identifying SATD. More advanced methods have been proposed using machine learning approaches \cite{maldonado2015detecting, da2017using, sala2021debthunter}, where classifiers like maximum entropy and Naive Bayes multinomial classifiers are trained to detect various types of TD. Recently, supervised deep learning methods have been introduced, demonstrating superior performance compared to rule-based and traditional machine learning approaches. Li \textit{et al.} \cite{li2023automatic} introduced a method called MT-Text-CNN, which utilizes a convolutional neural network combined with multi-task learning to detect SATD across multiple sources. Meanwhile, Yu \textit{et al.} \cite{yu2021using} proposed a method utilizing bidirectional long short-term memory (BiLSTM) networks with an attention mechanism and a balanced cross-entropy loss function to mitigate the imbalance problem in SATD identification. Besides, several approaches have utilized pretrained language models, achieving state-of-the-art performance in SATD-related tasks. Sharma \textit{et al.} \cite{sharma2022self} demonstrated that ALBERT \cite{lan2019albert} and RoBERTa \cite{liu2019roberta} significantly outperform traditional machine learning and CNN-based methods in identifying and classifying SATD categories in the R language. Furthermore, Sheikhari \textit{et al.} \cite{sheikhaei2024empirical} illustrated the superior performance of the LLM model Flan-T5 \cite{chung2024scaling} across three SATD-related tasks. \revised{Some methods incorporate corresponding source code to support SATD detection \citep{10.1145/3524842.3528469, zampetti2020automatically, russo2025leveraging}. For example, Russo \textit{et al.} \cite{10.1145/3524842.3528469} introduced WeakSATD to analyze C source code from Chromium projects, aiming to determine whether code blocks linked to SATD comments contain potential weaknesses. Meanwhile, VulSATD \cite{russo2025leveraging} leverages CodeBERT to jointly encode comment–code pairs in a multi-task setup, with separate output heads for identifying SATD and detecting code vulnerabilities.}

\section{Conclusions and Future Work} 
\label{sec:Conclusions}
In this work we created a pipeline to augment existing datasets for investigating SATD. 
By means of an empirical evaluation, we demonstrate that the pipeline improves both the quality and the scope of data used in SATD research by employing a selection strategy that identifies informative examples, thereby minimizing the manual labeling effort. Moreover, our study highlights the effectiveness of integrating additional source code context into SATD detection. This strategy improves both the accuracy and robustness of existing models. Our study is the first to provide a comprehensive analysis of how to efficiently leverage code context by examining different scopes surrounding comments. Additionally, we propose two effective 
methods to incorporate this contextual information and an ensemble approach combining multi-scope results to achieve superior performance in detecting SATD in code comments. Furthermore, we conducted extensive experiments using a large number of models, diverse in size, architecture, and knowledge domain, on a novel scenario--detecting technical debt in source code. These experiments provide valuable insights into the performance and adaptability of various models in this context.

We see this as just the first step in a program of research.  In our future research we can further improve the context for detecting technical debt in source code by incorporating additional information such as execution outputs, test coverage, and runtime metrics. These enhancements could lead to more accurate and comprehensive technical debt detection methods, which could benefit software maintenance and quality assurance processes.

\begin{acks}

    This paper has been partially supported by the MOSAICO project (Management, Orchestration and Supervision of AI-agent COmmunities for reliable AI in software engineering) that has received funding from the European Union under the Horizon Research and Innovation Action (Grant Agreement No. 101189664). The work has been partially supported by the EMELIOT national research project, which has been funded by the MUR under the PRIN 2020 program (Contract 2020W3A5FY). It has been also partially supported by the European Union--NextGenerationEU through the Italian Ministry of University and Research, Projects PRIN 2022 PNRR \emph{``FRINGE: context-aware FaiRness engineerING in complex software systEms''} grant n. P2022553SL. We acknowledge the Italian ``PRIN 2022'' project TRex-SE: \emph{``Trustworthy Recommenders for Software Engineers,''} grant n. 2022LKJWHC. Our research is also funded by Hanoi University of Science and Technology (HUST), Vietnam under project number T2023-PC-002. 
\end{acks}

\bibliographystyle{ACM-Reference-Format}
\bibliography{main}

\appendix

\end{document}